\documentclass[floatfix,aps,pre,showpacs]{revtex4}
\usepackage{latexsym} \usepackage{graphicx} \usepackage{amsmath}

\newcommand{\sig}[1]{\mathrm{Sign}#1} \def\argmin{\mbox{argmin}}
\def\ud{\mathrm{d}} \def\B{\beta}

\def\vecxmu{\vec{x}^\mu} \def\vecj{\vec{J}} \def\vecJ{\vec{J}}
\def\vecjcero{\vec{J^0}} \def\vecJcero{\vec{J^0}} 
\def\hatq{\hat{q}} \def\hatr{\hat{r}} \def\hatQ{\hat{Q}}
\def\hatR{\hat{R}} \def\rhoJ{\rho(J^0)}  \def\intrho{\int \ud J^0 \rho(J^0)}
\def\DNx{\mathrm{D}x\:} \def\arccot{\mbox{ArcCot}}
\def\neff{n_{\mbox{\tiny eff}}} \def\neffcero{n_{\mbox{\tiny
      eff}}^{0}} \def\neffth{n_{\mbox{\tiny eff}}^{th}}
 \newcommand{\eq}[1]{(\ref{#1})}
\newcommand{\norm}[1]{\| #1 \|_p} \def\Jth{J_{th}} \def\H{\mathcal{H}}
\def\Lcero{$\mbox{L}_0$ } \def\Luno{$\mbox{L}_1$ }
\def\Lp{$\mbox{L}_p$ }

\begin{document}
\title{Statistical mechanics of sparse generalization and model
  selection}

\author{Alejandro Lage-Castellanos } \affiliation{ Physics Faculty,
  University of Havana, La Habana, CP 10400, Cuba}
\affiliation{Institute for Scientific Interchange, Viale Settimio
  Severo 65, Villa Gualino, I-10133 Torino, Italy}

\author{Andrea Pagnani} \affiliation{Institute for Scientific
  Interchange, Viale Settimio Severo 65, Villa Gualino, I-10133
  Torino, Italy}

\author{Martin Weigt} \affiliation{Institute for Scientific
  Interchange, Viale Settimio Severo 65, Villa Gualino, I-10133
  Torino, Italy}

\date{\today}

%\begin{document}

%\maketitle

\begin{abstract}
One of the crucial tasks in many inference problems is the extraction
of sparse information out of a given number of high-dimensional
measurements. In machine learning, this is frequently achieved using,
as a penality term, the $L_p$ norm of the model parameters, with $p\leq
1$ for efficient dilution. Here we propose a statistical-mechanics
analysis of the problem in the setting of perceptron memorization and
generalization. Using a replica approach, we are able to evaluate the
relative performance of naive dilution (obtained by learning without
dilution, following by applying a threshold to the model parameters),
$L_1$ dilution (which is frequently used in convex optimization) and
$L_0$ dilution (which is optimal but computationally hard to
implement).  Whereas both $L_p$ diluted approaches clearly outperform
the naive approach, we find a small region where $L_0$ works almost
perfectly and strongly outperforms the simpler to implement $L_1$
dilution.
\end{abstract}
\pacs{02.50.Tt Inference methods, 05.20.-y Classical statistical mechanics}

\maketitle
\section{Introduction}
\label{sec:int}

The problem of extracting sparse information from high-dimensional data
is common to various fields of scientific data analysis: computational
biology, computer science, combinatorial chemistry, neuroscience, and
text processing are just a few examples (see 
\cite{Guyon_Elisseeff2003,Guyon_etal_2006} for a general introduction on 
the subject). Its importance becomes particularly evident in the analysis
of biological high-throughput experiments. To give an example, the number 
of gene probes analyzed simultaneously ranges from the order of tens of 
thousands in gene expression experiments (e.g. $\sim 30,000$ for human
DNA chips) to hundreds of thousands in the case of single-nucleotide
polymorphisms ($\sim 500,000$ for standard genotyping platforms). The
information about certain phenotypical traits is, however, expected to 
be contained in an {\it a priori} unknown, but small fraction (e.g. 
$<100$) of all measured probes. These probes may act in a combinatorial
way, making their one-by-one extraction impossible. As a further
complication, also the number of independent measurements rarely exceeds
the order of few hundreds. Therefore the problem of extracting information
from {\em few} high-dimensional data points has become a major
challenge in biological research. Both the extraction of features being
related to the phentypical traits (i.e. topological information) and
the construction of an explicit functional relation between the measured
values of these features and the phenotype are of enormous interest.

The literature about feature selection has so far been concentrated around 
two main strategies: (i) {\em wrapper} which utilizes learning to score 
signatures according to their predictive value, (ii) {\em filters} that 
fix the signature as a preprocessing step independent from the 
classification strategy used in the second step. In this work we will 
present a replica computation on a {\em wrapper} strategy which falls 
into the subclass of {\em embedded} methods where variable selection 
is performed in the training process of the classifier. More concretely, 
we will present an analytical teacher-student computation on the 
properties of a continuous diluted perceptron ({\em i.e.} a perceptron 
where a finite fraction of the coupling parameters are zero). Dilution 
will be introduced via an external field forcing the student to set as 
many variables as possible to zero. This external field will be coupled 
to the \Lp norm $||\vec J||_p = \sum_i |J_i|^p$ of the coupling vector 
of the student perceptron. For $p\leq 1$, the cusp-like singularity of
this function in zero actually sets a fraction of all model parameters
exactly to zero, as required for diluted inference. 

This strategy is not new, but so far, most of the more mathematically-minded 
studies in the context of linear regression and various non-linear models 
\cite{LASSO, wainwright, banerjee,koller,schmidt,meinshausen} have been 
concentrating (a) on the case $p=1$, which is the only case of a convex
\Lp norm with a cusp in zero, and therefore dilution can be achieved within
the framework of convex optimization (this case is well-known under the name
LASSO \cite{LASSO}); and (b) on the case of a large amount
of available data (our model parameter $\alpha$ would scale like $\ln N$
instead of being constant as in our setting), where mathematically
rigorous performance guarantees can be given.

It is, however, obvious, that the most efficient dilution should be
obtained for $p=0$, where non-zero parameters are penalized independently
of their non-zero value. The non-convexity of the $L_0$ norm introduces
computational complexity. Very few studies have been published so far for 
a binary sparse classifier: after a work of Malzahn \cite{Malzahn2000},
where the theoretical performance of a continuous and a ternary ({\em
i.e.} $\pm 1, 0$) perceptron are compared, the problem of the
inference of a classifier with discrete weights has been analyzed in
\cite{Japan,Martin08,Tria09}, where both a theoretical computation for 
the average case together with a message passing algorithm has been
proposed. Another way of attacking the problem has been recently
proposed by Kabashima in \cite{Kaba03,Uda,Kabashima2007}, where a
continuous perceptron whose variables are masked by boolean variables
mimicking dilution.

The article is organized as follows. In Sec. \ref{sec:gen} we the
describe the generalization problem, and the replica approach used for
its analytical description. In Secs. \ref{sec:Nondiluted} and 
\ref{sec:Diluted} we apply the general results of the replica trick to 
non-diluted generalization and \Lp diluted generalization respectively. 
The performance of the non-diluted, \Luno and \Lcero diluted generalizations 
are compared in Sec. \ref{sec:compare}. In Sec. \ref{sec:memorization} the
memorization problem is treated as a noise-dominated limiting case of
generalization, and at the end, the main results are reviewed and put
in context in the conclusions \ref{sec:conclusiones}. Three appendices
are added to clarify some technical aspects of the mathematical
derivations.

\section{Generalization and Replicas}
\label{sec:gen}

Two common problems in Machine Learning are the so-called Memorization
and Generalization problems. In either of them, a number of patterns
$\{\vecxmu, \mu\in(1\ldots M)\}$ are classified by labels $y^\mu$, and
one aims at memorizing or inferring a rule that reproduces the
given classification. We will study these problems, for the
perceptron with continuous weights.

Let us consider the case of $N$ binary variables $x_i=\pm 1$ defining
each pattern $\vecxmu$. We assume the existence of a {\it hidden}
relation among these variables and the labels $y^\mu=\pm 1$ of each
pattern:
\[y^\mu=\sigma^0(\vec{x}^\mu) \ .
\]
The function $\sigma^0(\vec{x})$ could be, e.g., the one relating the
activated/repressed expression states of genes $x_i$ with the presence
or absence $y=\pm 1$ of a disease, or with the expression of another
gene not contained in $\vec x$. Unfortunately, $\sigma^0(\vec{x})$ is
unknown and all we have in general is a set of $M$ experiments
$\{(y^\mu,\vecxmu), \mu\in(1\ldots M)\}$, linking patterns $\vecxmu$
to labels $y^\mu$. In supervised learning these experimental data are
used as a ``training set'' to infer the real relations among the
variables. As a first approximation, one could mimic the output
function $\sigma^0(\vec{x})$ as the sign of a linear combination,
\[
\sigma(\vecj,\vecxmu)=\sig(\vecj\cdot\vecxmu)\ ,
\]
where the $N$ weights $J_i$, also called couplings, are parameters to
be tuned in order to reproduce the experimental (training) data. Such
a function is called a perceptron. Here the weights $J$s are
allowed to take continuous real values.

The memorization \cite{Ga87,Gardner88} and generalization
\cite{Gy,Seung92} problems concern the question of inferring the
optimal values of the $J$s from the training data
$\{(y^\mu,\vecxmu)\}$. To this scope we define the training energy
(cost function)
\begin{equation}
E(\vecj)=\sum_\mu^M \Theta(-y^\mu \vecj\cdot\vecxmu ) \label{eq:E}
\end{equation}
counting the number of misclassified patterns when $\vecj$ is used to
reproduce the training data. The function $\Theta(\cdot)$ is the
Heaviside step function: $\Theta(x)=1$ if $x>0$, and zero
otherwise. Note that the function $E(\vecj)$ depends only on the
orientation of the vector and not on its length, {\em i.e} $E(\vecj) =
E(c \vecj)$ for all $c\neq 0$.

In general, the real unknown output function $\sigma^0(\vec{x})$ will
be a complex one, and attempts of reproducing it by a linear
perceptron may fail. This means that the training energy will
eventually become non zero if the number of training patterns is
sufficiently large. However, we will focus on the case of realizable
rules, this is, when the output function $\sigma^0(\vec{x})$ is
actually a perceptron, and there is always at least one set of weights
with zero energy.

The possibility of non-realizability will be accounted for as a random
noise affecting the output. In mathematical terms, the training
patterns are generated by
\begin{equation}
y^\mu=\sigma^0(\vec{x})= \sig(\vecjcero\cdot\vecxmu +
\eta^\mu) \label{eq:sigmacero}
\end{equation}
where the noise $\eta^\mu$ are i.i.d. Gaussian variables, with
variance $\gamma^2$, and the {\it hidden} perceptron parameters
$\vecjcero$ are the rule we are interested to ``discover''. We will
refer to $\vecjcero$ as the teacher, and to the free parameters of our
problem $\vecj$ as the student, since the latter pretends to reproduce
the patterns generated by the former.  Note that the training energy
\eq{eq:E} does not change when $\vecj$ is multiplied by a global
scalar factor. To cope with this invariance, we will look for student
vectors subject to the spherical constraint $\vecJ\cdot\vecj=N$.

In the zero noise limit ($\gamma\rightarrow 0$), there will be at
least one student capable of correctly classifying any amount of
training data, namely $\vecJ=\vecJcero$. Upon increasing the noise
level ($\gamma> 0$), the correlation between the patterns and the
teacher becomes shadowed by the noise, and the student will need a
larger amount of patterns to learn the teacher. If the noise dominates
completely $\gamma\to \infty$, there is no information left in the
training data about the teacher's structure, and the student will
memorize all patterns up to a critical threshold above which starts
to fail..

In the case of a feasible rule, the number of perfect solutions for
the student ($E(\vecj)=0$) is generally large. The entropy of the
space of perfect solutions is a decreasing function of the number $M$
of training patterns, since every new pattern imposes a constraint to
the student. We can further restrict this space by looking at diluted
solutions inside the space of perfect students. A general dilution term
can be added to the training energy to form the following Hamiltonian
\begin{equation}
\B\H(\vecj)=\B E(\vecJ) + h \norm{\vecJ} \label{eq:Ham}
\end{equation}
where $\norm{\vecJ}=\sum_i |J_i|^p$ is the $L_p$ norm of the
student. The dilution field $h$ will be used to force dilution, and
non-diluted generalization correspond to $h=0$. Among the different
choices of $p$, the case $p=1$ corresponds to the \Luno norm
$\|\vecj\|_1=\sum_i^N |J_i|$ used in the celebrated Tibshirani's paper
\cite{LASSO}, while $p=0$ corresponds to the \Lcero norm
$\|\vecj\|_0=\sum_i^N (1-\delta_{J_i})$, where $\delta_{J}$ is the
Kronecker delta. A particular feature of the $L_p$-norm is that, for
$p\leq 1$, it sets a finite fraction of the model parameters exactly
to zero, whereas it is convex for $p\geq 1$. The only parameter common
to these two ranges is $p=1$, explaining the popularity of the
$L_1$-norm for convex optimization approaches.

In the following we apply the replica trick to compute the volume of
the space of solutions \cite{Ga87,BrEn}, as well as other relevant
quantities (order parameters) for the generalization problem.
 
\subsection{Replica calculation}
\label{sec:replicas}

Let us consider the space of optimal solutions for the supervised
learning of a realizable rule. The standard situation would be that a
training set $\{(y^\mu,\vecxmu)\}$ of $M$ experiments is presented to
be classified by a linear perceptron with $N$ continuous weights
$J_i$. The number of patterns relative to the amount of variables,
$\alpha=M/N$, will play an essential role as a control parameter. We
define the Gibbs measure for the student vector $\vecJ$ as
\[ P_{\mbox{\tiny Gibbs}}(\vecJ)=\frac{1}{Z(\beta,h)}
e^{-\beta E(\vecj) -h \norm{\vecJ} }
\]
It depends on the inverse temperature $\beta$, and the dilution field
$h$. In the $\B\rightarrow \infty$ limit, the partition function
\[
Z(\beta,h)\:=\:\:\int \prod_{i=1}^N \ud J_i \:\:\exp\left(-\beta
E(\vecj) -h \norm{\vecJ} \right)
\]
contains only terms of minimal training energy. So, by computing $Z$
we can obtain the properties of the desired space. Although not
explicit indicated, the integration should is over the sphere
$\vecJ\cdot\vecJ=N$ to remove the scale invariance in the energy term
in Eq.~\eq{eq:Ham}.

In the partition function above, the degrees of freedom are the $N$
couplings $J_i$, while the $\vecxmu$ and $y^\mu$ present in the
Hamiltonian is the so called quenched disorder. As we care about the
properties of the solutions in the typical case, we will have to
average over these quenched variables. In particular, the $\vecxmu$
will be i.i.d. random variables in $\{\pm 1\}^N$, while the labels
$y^\mu$ are generated from the hidden structure of the couplings by
equation \eq{eq:sigmacero}. The teacher weights $\vecjcero$ too are
i.i.d.  i random variables distributed as:
\begin{equation}
\rhoJ=(1-\neffcero) \delta_{J^0} + \neffcero
\rho'(J^0) \ . \label{eq:rhoJ}
\end{equation}
The first term introduces the sparsity of the teacher, and the second
term contains all non zero couplings. Later we will use the letter $t$
to refer to the variance of this distribution. The effective fraction
of couplings $\neffcero=\frac{N_{J\neq 0}}{N}$, is the relative amount
of non-zero couplings, and sparse models correspond to small
$\neffcero\ll1$. The fact that $\vecjcero$ is involved directly in the
computation will allow us to compare the student vector $\vecJ$ to it.

The free energy $f=-\frac{1}{\B} \log Z$ is the relevant thermodynamic
quantity, and the one that should be averaged over the quenched
disorder. However, the direct integration over $\vecxmu$ and
$\vecjcero$ in $\overline{\log Z}$ is out of reach. To work around
this obstacle, we use the replica trick \cite{SGTB}, which consist of
using the known property
\begin{equation}
\log Z= \lim_{n\to 0} \frac{Z^n-1}{n} \label{eq:Z}
\end{equation}
to average over $Z^n$, instead of $\log Z$, and sending $n$ to zero
afterwards. Note that $Z^n$ is the partition function of a $n$-times
replicated system, if $n$ is integer, which is the origin of the
method's name. In our case the averaged and replicated partition
function would be
\begin{eqnarray}
\overline{Z^n}&=&2^{-M N} \sum_{x^\mu_i=\pm 1}\; \int \prod_{\mu=1}^M
\mathrm{D_\gamma}\eta_\mu \int \prod_{i=1}^N \ud J^0_i
\:\prod_{i=1}^N \rho(J_i^0) \int \prod_{a=1}^n \prod_{i=1}^N \ud
J^a_i \label{eq:Zn} \\ &&\phantom{2^{-M N} \sum_{x^\mu_i}\;} \exp
\left\lbrace -\B \sum_{a=1}^n \sum_{\mu=1}^M \Theta\left(
-\left[\sum_{i=1}^{N} J_i^0 x_i^\mu +\eta_\mu\right]
\left[\sum_{i=1}^{N} J_i^a x_i^\mu \right]\right) - h \;\sum_{a=1}^{n}
\norm{\vecJ^a} \right\rbrace \nonumber
\end{eqnarray}
where $\mathrm{D_\gamma}\eta_\mu$ stands for the Gaussian
distributions of the noise variable $\eta_\mu$, with variance
$\gamma^2$,
\[\mathrm{D_\gamma}\eta_\mu=\frac{e^{-\frac{{\eta_\mu}^2}{2 \gamma^2}}}{ 
\gamma \sqrt{2 \pi}} \ud \eta_\mu
\]
This notation will be used throughout the paper, and if the subindex
$\gamma$ is omitted, it refers to $\gamma=1$.

After some standard steps detailed in appendix \ref{ap:Replica}, the
replica symmetric estimate of the free energy is obtained as
\begin{equation}
 -\B \overline{f}= {\rm extr}_{q, \hatq, r, \hatr, \lambda} 
\left\{ -r\hatr+\frac{1}{2}q \hatq -\lambda + G_J+\alpha\:
 G_X \right\} \nonumber% \label{eq:freeenergy}
\end{equation}
The order parameters $q, \hatq, r, \hatr$ and $\lambda$ were
introduced via Dirac-delta functions in the replica calculation. In
particular $q=N^{-1}<\vecJ^a \cdot\vecJ^b>$ is the overlap between two
(independent) students solutions. The notation $<\cdot>$ stands for
the expectation value w.r.t the Gibbs measure. Note that $0\leq
q\leq1$, it will be $1$ when the Gibbs measure is condensed in a
single $\vecJ$, and it will be smaller than one when the measure is
more spread. The parameter $r=N^{-1}<\vecJ \cdot\vecjcero>$ is the
overlap between the student vectors and the teacher, and will be
crucial in our understanding of the performance of generalization. The
parameters $\hatq$, $\hatr$, and $\lambda$ are the corresponding
associated Fourier variables (to represent the deltas introduced in
the replica calculation). The last one, $\lambda$, corresponds to the
spherical constraint $\vecJ\cdot \vecJ =N$.

The terms $G_J$ and $G_X$ are given by
\begin{eqnarray}
G_J & =&\int \DNx \intrho \log\int \ud J
\exp\left(-(\frac{\hatq}{2}-\lambda)J^2 -h \norm{J} +(\hatr J^0-
\sqrt{\hatq} x) J\right) \ ,\label{eq:GJGX}\\ G_X & = & 2 \int \DNx
H\left(\frac{ x r}{\sqrt{q \gamma^2+q t-r^2)} }\right)
\log\left((e^{-\B}-1) H(-\sqrt{\frac{q}{1-q}}x)+1 \right) \nonumber
\end{eqnarray}
with $H(x)=\int_x^\infty \frac{\ud y}{\sqrt{2 \pi}}e^{-y^2/2}$. From
the replica calculation, the term $G_J$ can be interpreted as the
effective free energy of a single $J$. The inner term
$\mathcal{Z}_J(J^0,x)=\int \ud J
\exp\left(-(\frac{\hatq}{2}-\lambda)J^2 -h \norm{J} +(\hatr J^0-
\sqrt{\hatq} x) J\right)$ plays the role of a single $J$ partition
function, while the term $\log \mathcal{Z}_J$ corresponds to its free
energy. The dependence of $\mathcal{Z}_J(J^0,x)$ on $J^0$ and $x$ is
conditioning the free energy of the single $J$ to the different values
$J^0$ of the corresponding element in the teacher vector, and to the
effective ``noise'' from the realization of the training patterns
$\vecxmu$. So the integration over $J^0$ and $x$ gives the average
effective free energy of a single $J$.

This interpretation of $G_J$ allows also for formulating the following
joint probability distribution of $x, J^0$ and $J$
\begin{equation}
P(x,J^0,J)=\frac{e^{-\frac{x^2}{2} } }{\sqrt{2 \pi}} \rhoJ \frac{
  e^{-(\frac{\hatq}{2}-\lambda)J^2 -h \norm{J} +(\hatr J^0-
    \sqrt{\hatq} x) J}}{\int \ud J e^{-(\frac{\hatq}{2}-\lambda)J^2 -h
    \norm{J} +(\hatr J^0- \sqrt{\hatq} x) J}} \label{eq:Pdil}
\end{equation}
such that any expectation value of a generic function $g(x,J^0,J)$ can
be found as
\[
\mathbf{E}[g(x,J^0,J)]=\int \ud x \int \ud J^0 \int \ud J\: g(x,J^0,J)
P(x,J^0,J)
\]

The limit $\B\to \infty$ is trivial in Eq. \eq{eq:GJGX}. It
concentrates the Gibbs measure onto the subspace of students with
minimum training energy (error), and in the case of a feasible rule to
the perfect solutions $E(\vecj)=0$. The actual values of the
variational parameters are determined by the saddle-point condition
for the free energy $\nabla_{q,r,\ldots} f =0$.  With all the previous
definitions, at zero temperature ($\B\to \infty$) this condition is
given by
\begin{eqnarray}
\hatq &=& \frac{r \hatr }{q} + \frac{\alpha \sqrt{2} }{\sqrt{ \pi}
  \sqrt{(1-q) q}} \int \DNx H\left(\frac{x r \sqrt{1-q}}{\sqrt{q
    \gamma^2+q t-r^2}} \right) \frac{x}{H(\sqrt{q} x)} \nonumber
\\ \hatr &=& \frac{-2 \alpha}{\sqrt{2 \pi} \sqrt{q \gamma^2+q t-r^2} }
\int \DNx \:\:x\: \log H\left( \sqrt{\frac{q}{1-q}}
\sqrt{1-\frac{r^2}{q \gamma^2+q t}} x\right) \nonumber \\ q&=& 1 +
\frac{1}{\sqrt{\hatq}}\mathbf{E}[x J] \label{eq:fixedpoint}\\ r&=&
\mathbf{E}[J^0 J] \nonumber\\ 1&=& \mathbf{E}[J^2] \nonumber
\end{eqnarray}

This set of equations has to be solved numerically for each
$\alpha=M/N$ and each dilution field $h$. The resulting values of the
variational parameters $q, r, \hatq, \hatr$ and $\lambda$ are used to
describe the solution space. For instance the generalization error,
i.e. the probability that a new pattern (independently generated from
those used for training) is misclassified by the student, depends only
on the overlap between teacher and student $r$ (see \cite{Seung92})
\begin{equation}
\epsilon = \frac{1}{\pi} \arccos
\frac{r}{\sqrt{t}} \label{eq:generror}
\end{equation}
The square root of the variance of the teacher $t=\intrho {J^0}^2$ is
required because the teacher is not necessarily normalized to unity.

The solution of the fixed point equations can also be used to
construct the Precision {\it vs} Recall curve, which is a standard
check for a classifier. In the case of model selection we can use the
information given by the student solution $\vecJ$ to classify the
couplings as relevant $J_i>\Jth$ or not relevant $J_i<\Jth$, where
$\Jth$ is a sensibility parameter. This means that we will disregard
all inferred $J_i$ which are not strong enough. With the joint
probability distribution \eq{eq:Pdil} we can compute the probability
of having any of the following situations
\begin{displaymath}
% use packages: array
\begin{array}{ll}
\mbox{True Positive TP}\quad & J_i\neq0\:\:\: J_i^0\neq0
\\ \mbox{False Positive FP}\quad & J_i\neq0\:\:\: J_i^0=0
\\ \mbox{True Negative TN}\quad & J_i=0\:\:\: J_i^0=0 \\ \mbox{False
  Negative FN}\quad & J_i=0\:\:\: J_i^0\neq0
\end{array}
\end{displaymath}
For instance the probability of having a true positive (TP) is
$P_{TP}=\mathbf{E}[\Theta(|J|-\Jth) (1-\delta_{J^0})]$.

The recall (sensitivity) and the precision (specifity) are defined as
follows
\begin{equation} RC=\frac{P_{TP}}{P_{TP}+P_{FN}}=\frac{P_{TP}}{n^0_{\mbox{eff}}} \qquad  PR=\frac{P_{TP}}{P_{TP}+P_{FP}}=\frac{P_{TP}}{n^{th}_{\rm eff}} \label{eq:PRRC}
\end{equation}
where $\neffcero$ is the real sparsity of the teacher (see
\eq{eq:rhoJ}) and $\neffth=\mathbf{E}[|J|>\Jth]$ is the dilution of
the student when the threshold value for a relevant coupling is
$\Jth$. Note that both the recall and the precision depend on $\Jth$,
as well as on the variational parameters $q,r,\hatq, \hatr$ and $
\lambda$ that solve the fixed point equations \eq{eq:fixedpoint}. The
PR-RC curve is the parametric curve $RC(J_{th})$ vs $PR(J_{th})$: 
the closer we can get to $RC=1$ and $PR=1$, the better the student
perceptron has understood the topological structure of the teacher.

\section{Non-diluted generalization}
\label{sec:Nondiluted}

To avoid confusion we will call {\it sparse} the case of teachers with
many trivial couplings $J_i^0=0$, while the term diluted will be saved
for the generalization method (non-diluted/diluted). The replica
calculation hitherto developed is general in a set of aspects. First,
the teacher distribution \eq{eq:rhoJ} can be of any kind, including a
non sparse teacher $\neffcero=1$, although we will focus on the case
of sparse models. Second, the possibility of a non-diluted
generalization can be accounted by setting the dilution field $h=0$,
and for $h\neq0$ different choices of regularization are possible. In
this paper we will show the results for \Lcero and \Luno. For each of
these cases (non-diluted, \Lcero and \Luno) the replica calculation
has it's particularities, which we present hereafter.

The simplest case is the non-diluted generalization ($h=0$), as some
of the equations simplify considerably, being equivalent to those in
\cite{Seung92}. The absence of the dilution term in \eq{eq:GJGX} makes
the expression integrable, such that
\[
 G_J = \frac{\hatr^2 t +\hatq}{2(\hatq-2
   \lambda)}-\frac{1}{2}\log(\hatq/2 -\lambda)
\]
The first two fixed point equations in \eq{eq:fixedpoint} do not
change, while the last three can be reduced to two algebraic equations
without $\lambda$,
\begin{eqnarray}
q &=& (\hatr^2 t + \hatq)(1-q)^2 \label{eq:fixedpointalgebric} \\ r
&=& \hatr t (1-q) \nonumber
\end{eqnarray}
The value of $\lambda$ can be recovered using $(1-q)(\hatq-2
\lambda)=1$. The fixed point equations can be solved numerically for
evaluating the generalization error \eq{eq:generror} as well as the
PR-RC curve. The calculation of the expectation values using
\eq{eq:Pdil} is also simplified since
\begin{equation}
 P(J^0,J)=\frac{\hatq-2 \lambda}{\sqrt{2 \pi (2\hatq-2\lambda)}} \rhoJ
 \exp{-\frac{((\hatq-2\lambda)J-\hatr J^0)^2}{2 (2 \hatq
     -2\lambda)}} \label{eq:Pnondil}
\end{equation}
and the terms involved in recall and precision \eq{eq:PRRC} are easier
to obtain.
\begin{figure}[htb]
        \begin{center}              
             \includegraphics[scale=0.4, angle=-90]{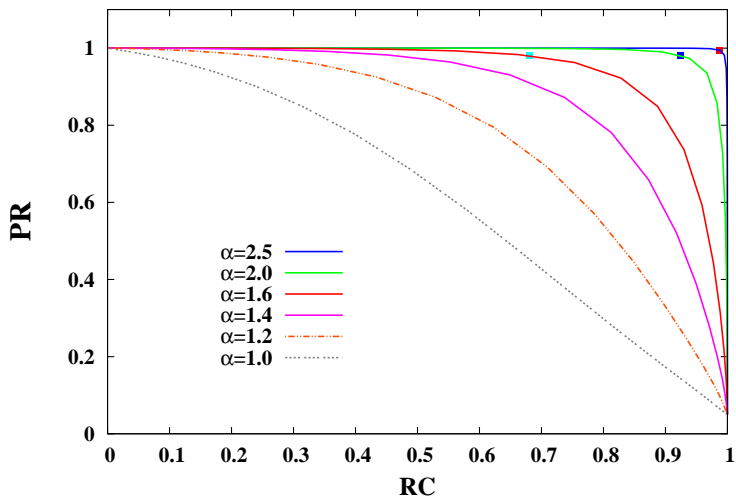}
            \end{center} 
	\caption{The Precision-Recall curve for the non-diluted
          generalization for different values of $\alpha$. For growing
          values of the amount of training data $\alpha$, the curves
          approach the $PR\equiv1$ line, meaning that the student
          solution is doing an almost perfect model selection, for a
          certain value of the sensitivity threshold $\Jth$.}
\label{fig:ROCNonDil}
 \end{figure}

Let us take as a toy example the case of a sparse teacher with only
$\neffcero=5\%$ non-zero couplings. We set the noise to $\gamma=0$,
such that there is always a perfect student solution. In particular,
we will use a discrete teacher $J^0\in\{-1,0,1\}$
\begin{equation}
\rhoJ=(1-0.05) \delta_{J^0} + \frac{0.05}{2}\left( \delta_{J^0,-1}
+\delta_{J^0,1} \right) \label{eq:toyrho}
\end{equation}
With such a simple structure it happens to be the case that teacher's
dilution and variance are both equal $\neffcero=t=0.05$.

The solution of the fixed point equations (\eq{eq:fixedpoint} and
\eq{eq:fixedpointalgebric}) is found numerically for different values
of the amount of training data $\alpha$. For each $\alpha$, the
different PR-RC curves are shown in Figure \ref{fig:ROCNonDil}. It is
clear from the figure that for sufficiently large $\alpha$, for
instance $\alpha\geq 2.0$ in this example, the generalization is
capable of a good classification of the couplings $J$, achieving both
high precision and high recall, i.e. a good performance in model
selection. This is seen in the figure as a curve that approaches the
$PR\equiv 1$ line.
 \begin{figure}[htb]
        \begin{center}              
             \includegraphics[scale=0.4, angle=-90]{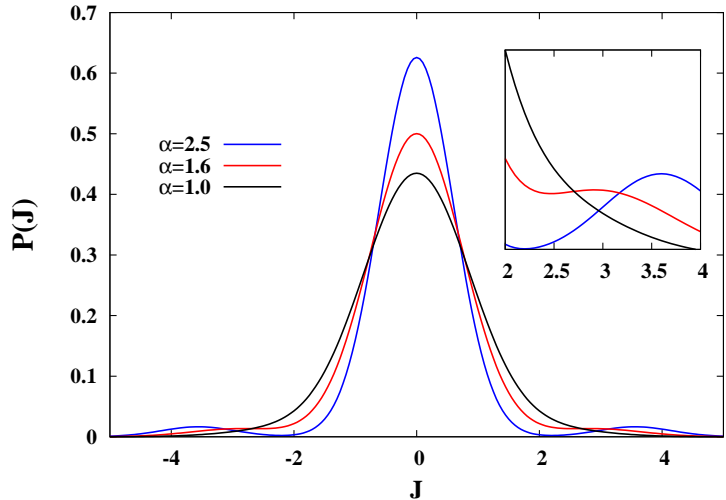}
            \end{center} 
	\caption{The statistical distribution of the students
          couplings $J$ for three different values of $\alpha$. When
          enough training data is given ($\alpha =2.5$ in this
          figure), the distribution $P(J)$ can be recognized as the
          superposition of Gaussian distributions located around the
          discrete (and rescaled) values of the teacher's
          couplings. In such a case, setting to $0$ all those
          student's couplings $J_i$ that are around $0$ results in a
          nearly perfect model reconstruction. As less information is
          used for training, the Gaussians overlap, and any threshold
          for the relevance of a coupling $\Jth$ will misclassify some
          couplings, resulting in a worse model selection.}
\label{fig:PJ}
 \end{figure}

It is no surprise that more training data results in better model
selection. However, we can gain some information about how the
solution approaches perfect model selection by looking at the
statistical distributions of the student's $J$s. Figure \ref{fig:PJ}
shows how $P(J)=\int \ud J^0 P(J^0,J)$ (Eq. \eq{eq:Pnondil})
concentrates around the discrete (and rescaled) values of $J^0$ with a
set of Gaussians that have neglectable overlap for large values of
$\alpha$. Above a critical $\alpha\simeq 1.6$ (in this example) we can
start to discriminate the $J$s from different Gaussians because local
minima in $P(J)$ emerge. It is expected that above this point a
reasonable value for $\Jth$ is the one satisfying
\[ \frac{\partial P(J)}{\partial J} =0 \quad 
\frac{\partial^2 P(J)}{\partial J^2} >0 \ .
\]
This choice for $\Jth$ leads to the recall and precision, which can be
seen in Figure \ref{fig:ROCNonDil} marked by the square symbols. For
$\alpha=2.5$ the Gaussians in $P(J)$ are almost perfectly
distinguishable.  The optimal choice for $\Jth$ has a precision and a
recall near one $(RC\cong0.991,PR\cong0.996)$, meaning that
generalization is achieving an almost perfect model selection.

\section{Diluted generalization}
\label{sec:Diluted}

At zero temperature (infinite $\B$) and $h=0$, the Gibbs measure gives
the same probability to all perfect student solutions, since they have
the same energy $E(\vecJ)=0$, while suppressing completely
positive-cost students. Working at zero temperature, a dilution field
$h>0$ gives the chance to impose a different measure over the set of
perfect solutions. This measure favors the students with the lowest
values of $\norm{\vecJ}$, and in the limit of $h\to\infty$, it
concentrates in the perfect solution with the highest \Lp dilution. We
will now study the properties of the subset of perfect students with
the smallest \Lp norm.

Unlike the trivial $\B\rightarrow\infty$ limit (see Eq. \eq{eq:GJGX}),
the large-dilution limit has to be taken carefully as some parameters
diverge. For large dilution field $h\to\infty$ we have $q\rightarrow
1$, meaning that different students are very close to each other, and
in the limit $h= \infty$ there is only one student which is at the
same time zero-cost ($E(\vecJ)=0$) and maximally diluted. As can be
seen from the fixed point equations \eq{eq:fixedpoint}, when $q$ tends
to 1, the order parameters $\hatq$, $\hatr$ and $\lambda$ diverge. The
scaling behavior of these variables is the following
\begin{equation}
 \begin{array}{rcllrcl}
  (1-q) &\simeq& \frac{Q}{ h} &\quad& \hatr&\simeq&\hatR h
   \\ \hatq&\simeq&\hatQ h^2 &\quad& \frac{\hatq}{2}-\lambda &\simeq&
   \frac{K}{2} h
 \end{array}\label{eq:scalingh}
\end{equation}
In terms of these new variables, the fixed point equations for $\hatq$
and $\hatr$ \eq{eq:fixedpoint} become
\begin{eqnarray}
 \hatQ&=&\frac{\alpha}{\pi Q^2} \left[ \arccot
   \frac{r}{\sqrt{\gamma^2+t-r^2}} - \frac{r
     \sqrt{\gamma^2+t-r^2}}{\gamma^2+t}\right] \nonumber
 \\ \hatR&=&\frac{\alpha \sqrt{\gamma^2+t-r^2}}{Q\pi
   (\gamma^2+t)} \label{eq:fphatQhatR}
\end{eqnarray}
and do not depend on the dilution $p$. Furthermore, this scaling makes
the exponent of the exponential term inside $P[x,J^0,J]$
(Eq. \eq{eq:Pdil}) proportional to $h$. So, in the remaining three
equations
\begin{eqnarray}
Q &=& \frac{1}{\sqrt{\hatQ}}\mathbf{E}[x J] \nonumber\\ r&=&
\mathbf{E}[J^0 J] \label{eq:fpQrK}\\ 1&=& \mathbf{E}[J^2] \nonumber
\end{eqnarray}
the expectation values for $h\to \infty$ are dominated by the largest
values of the exponent in \eq{eq:Pdil}. The specific details of this
saddle-point calculation depend on the actual dilution $\norm{\vecJ}$.

Among all possible values of $p$, the cases $p=1$ and $p=0$ are
special for both their meaning and their simplicity in the
calculations. The \Luno norm is extremely popular in machine learning
because it maintains the convexity of a convex cost function while
forcing sparse solutions \cite{LASSO}. On the other hand, \Lcero lacks
completely of the convexity preserving property (it is not even
continuous), and therefore is not a suitable penalization for convex
optimization. However, the \Lcero norm is optimal in the sense that it
does not deform the Hamiltonian beyond penalizing non-zero
couplings. We will compare the dilution achieved by the \Luno approach
with the largest possible dilution (the one obtained using \Lcero),
and give a qualitative description of this widely used
regularization. Another simple and common choice for the penalization
is $p=2$, but in our model setting it is meaningless since the student
is constrained to the sphere and therefore has a fixed
$\|\vecJ\|_2=N$.

\subsection{\Luno dilution}

We first discuss the case of \Luno-regularization
$\|\vecJ\|_1=\sum_{i=1}^N|J_i|$. For $h\to \infty$ the expectation
value of an arbitrary function $g(x,J^0,J)$ is given by \[
  \mathbf{E}[g(x,J^0,J)]=\int \DNx \intrho \times \left\{\begin{array}{lll}
  g(x,J^0,0) & \:\: & |\hatR J^0 -\sqrt{\hatQ} x|<1
  \\ g(x,J^0,\frac{\hatR J^0 -\sqrt{\hatQ} x -\sig(\hatR J^0
    -\sqrt{\hatQ} x)}{ K}) & \:\: & \mbox{ otherwise.}
\end{array} \right. 
\]
The derivation of this expectation value is shown in appendix
\ref{ap:hlimit}. As already mentioned, one of the virtues of the
\Luno-regularization is that it forces the solution to be diluted by
setting a fraction of the couplings exactly to zero. This fact becomes
evident in the previous equation.

Using the scaling behavior \eq{eq:scalingh}, and calling
$S(J^0,x)=\sig(\hatR J^0 -\sqrt{\hatQ} x)$, and defining the functional
\[L[\cdot]=\int \DNx [\cdot] \Theta(|\hatR J^0 -\sqrt{\hatQ} x|-1)\:\mbox{,}\] 
the resulting fixed-point equations in the $h\to \infty$ limit
are \eq{eq:fphatQhatR} and
\begin{eqnarray}
Q &=&\frac{1}{K} \intrho L[1] \label{eq:fixedpointL1} \\ r
&=&\frac{1}{K} \intrho \left[ \hatR {J^0}^2 L[1] -\sqrt{\hatQ} J^0
  L[x] -J^0 L[S(J^0,x)] \right] \nonumber \\ K &=& (\hatQ +1 )Q + r
\hatR + \frac{1}{K}\intrho \left[\sqrt{\hatQ} L[x S(J^0,x)] - \hatR
  J^0 L[S(J^0,x)]\right]\nonumber
\end{eqnarray}

Note that the original parameter $q$ is no longer present, since it is
$1$, but the overlap between teacher and student, $r$, is still a non
trivial order parameter.

\subsection{\Lcero dilution}
For the \Lcero-regularization the dilution term in the Hamiltonian
\eq{eq:Ham} is $\|\vecJ\|_0=\sum_i^N(1-\delta_{J_i})$, punishing only
the fact that a given $J_i$ is non zero, but otherwise making no
distinction between different non-zero $J$-values. In a strict
mathematical sense, introducing the \Lcero norm in the Hamiltonian is
meaningless, since a finite single-point discontinuity cannot alter
the integration over the continuous range of $J$-values in the
partition function. So the \Lcero norm can only be understood as the
limiting case $p\to +0$ of a family of continuous functions (see
appendix \ref{ap:plimit}).

Using a similar approach as the one presented in appendix
\ref{ap:hlimit} for \Luno, the expectation value of an arbitrary
function $g(x,J^0,J)$ in the $h\to \infty$ limit reads
{\small\begin{equation} \mathbf{E}[g(x,J^0,J)]=\int \DNx \intrho
\left\{\begin{array}{lll} g(x,J^0,0) & \:\: & \frac{|\hatR J^0
-\sqrt{\hatQ} x|}{\sqrt{2 K}} <1 \\ g(x,J^0,\frac{\hatR J^0
-\sqrt{\hatQ} x}{ K}) & \:\: & \mbox{ otherwise}
\end{array} \right. \label{eq:ExpectedL0}
       \end{equation}}

The fixed-point equations for $\hatQ$ and $\hatR$ are exactly the same
as for \Luno-dilution (eq. \eq{eq:fphatQhatR}), while the other three
order parameters are now given by
\begin{eqnarray}
Q &=&\frac{1}{K} \intrho \left[ M[1] -\frac{1}{\sqrt{\hatQ}} \tilde M[x
  S(J^0,x)]\right] \nonumber\\ r &=&\frac{1}{K} \intrho \left[ \hatR
  {J^0}^2 M[1] -\sqrt{\hatQ} J^0 M[x] \right] \label{eq:fixedpointL0}
\\ K &=& r \hatR + Q \hatQ\nonumber
\end{eqnarray}
where $M[\cdot]=\int \DNx \:[\cdot]\: \theta(\frac{|\hatR J^0
  -\sqrt{\hatQ} x|}{\sqrt{2 K}} -1 )$, and $\tilde M[x
S(J^0,x)]$ is defined as
\[\tilde M[x S(J^0,x)] = -\frac{2 \sqrt{2 K}}{\sqrt{2 \pi}} 
\exp{-\frac{\hatR^2 {J^0}^2+2 K}{2 \hatQ}} \cosh{\frac{\hatR J^0 
\sqrt{2 K}}{\hatQ}} \ .
\]

\subsection{Dilution, recall and precision}

The numerical solution of the fixed-point equations for the \Luno and
\Lcero dilutions gives us the overlap $r$ between the teacher and the
student. The generalization error is obtained using
Eq. \eq{eq:generror}.

The most striking effect of the norms is the emergence of an extensive
number of couplings that are exactly zero (see $P(J)$ in
appendices). The fraction of non-zero couplings is the effective
dilution $\neff$ achieved by the student, and it is obtained as
\begin{equation}
% use packages: array
 \neff \:=\mathbf{E}[|J|>0] =\left\{\begin{array}{llll} \intrho L[1]
 \:= Q K &\:\:\: L_1 \mbox{ norm} \\ \intrho M[1] &\:\:\: L_0 \mbox{
   norm}
\end{array}\right. \label{eq:neff}
\end{equation}
It is expected (and numerically observed) that for large values of
$\alpha$ the effective dilution $\neff$ converges to the real dilution
of the teacher $\neffcero$.

Along the same lines developed for the non-diluted case, we can
further restrict the set of non-zero couplings by setting a threshold
for relevant couplings. In other words, we interpret as non-relevant
all those couplings that are not strong enough, $J_i<|\Jth|$. In this
case, the fraction of relevant couplings equals
\[ \neffth \:=\mathbf{E}[\Theta(|J|-\Jth)]
\]
and can be used to calculate the precision according to
Eq. \eq{eq:PRRC}. The other terms appearing in recall and
precision are also computed using the expectation value
$\mathbf{E}[\cdot]$ for each dilution scheme. For instance, the
probability of having a false positive is given by
$P_{FP}=\mathbf{E}[\Theta(|J|-\Jth) \delta_{J^0}]$.

\section{How well does dilution work?}
\label{sec:compare}
The discussed mathematical machinery can shed some light on this
question. To see the differences between diluted and non-diluted
generalization, and its performance in sparse model selection, let us
use the same toy example used for the non-diluted case, with a teacher
of dilution $\neffcero=5\%$ and discrete values $J_i^0\in\{-1,0,1\}$,
see Eq. \eq{eq:toyrho}.
\begin{figure}[htb]
        \begin{center}              
             \includegraphics[scale=0.8]{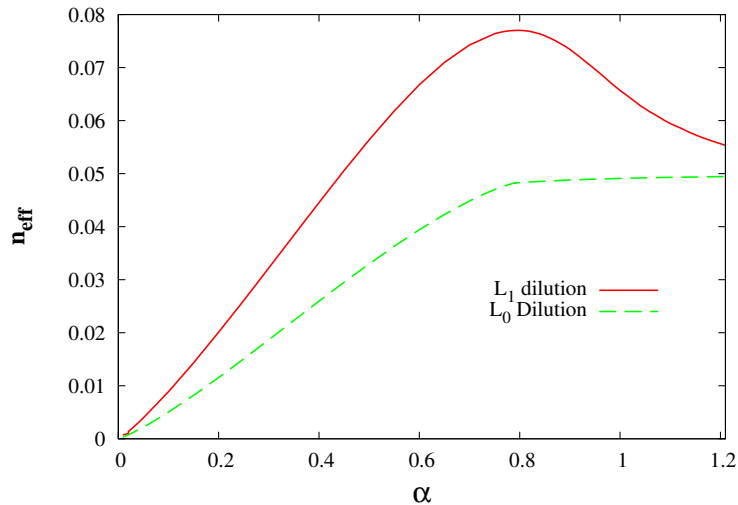}
            \end{center} 
	\caption{The dilution $\neff(\alpha)$ achieved by the \Luno
          and \Lcero dilutions, as a function of the amount of
          training patterns $\alpha$. The \Lcero regularization
          approaches the dilution of the teacher $\neffcero=0.05$
          from below, including non zero couplings only when strictly
          required to correctly classify the training data. The \Luno
          dilution is not that efficient, and for $\alpha > 0.45$ it
          uses more non-zero couplings than actually needed.}
\label{fig:Genneff005}
 \end{figure}

The functions $\neff(\alpha)$ for the \Luno and \Lcero dilutions are
presented in Fig. \ref{fig:Genneff005}. It can be seen that
\Lcero-diluted generalization goes monotonously from below to the
correct value $\neff=0.05$, in a somehow Ocams-optimal way. In other
words, \Lcero dilution adds non-zero couplings just when strictly
required by the empirical (training) evidence. The \Luno norm isn't
that effective. It is an interesting result that, for a certain range
in $\alpha$, the \Luno optimal solution requires more non-zero
couplings ($\neff>\neffcero=0.05$) than actually present in the
teacher. This overshooting is the cost we pay for deforming the
Hamiltonian by the \Luno-penalization of large couplings. Unlike \Lcero
regularization, \Luno approaches the correct dilution from above, not
from below.
\begin{figure}[htb]
        \begin{center}              
             \includegraphics[scale=0.8]{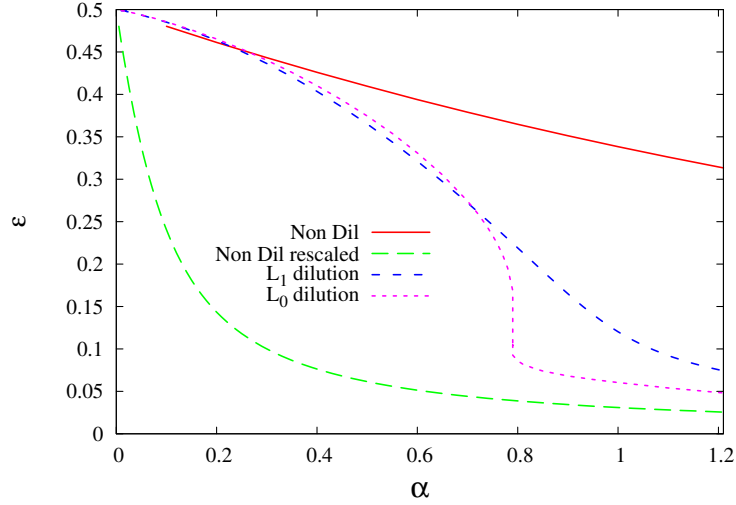}
            \end{center} 
	\caption{The generalization error at different $\alpha$. The
          two curves at the center are the generalization error
          achieved using \Luno and \Lcero dilution. While the \Luno
          error decreases smoothly with the training data, the one
          corresponding to \Lcero undergoes a sudden drop near
          $\alpha=0.8$. This is a consequence of a sudden move to
          almost perfect model selection, where the set of non-zero
          interactions has been identified with very good
          precision. Both, the superior and lower curves, correspond
          to the non-diluted generalization, where the lower one has
          been plotted with rescaled $x$-axis,
          $\epsilon(\alpha/\neffcero)$. If the student could know from
          the beginning which are the non-zero couplings, it could use
          all the training data to tune the values of these couplings,
          resulting in a huge reduction of the generalization error. }
\label{fig:generror}
 \end{figure}

One could be tempted to call the change of slope of \Lcero near
$\alpha=0.8$ in Fig. \ref{fig:Genneff005} a transition to perfect
student solution, but it is not. The generalization error in
Fig. \ref{fig:generror} shows that errors persist also for larger
$\alpha$. On the other hand, while the \Luno norm goes smoothly to
$\epsilon=0$, the \Lcero undergoes an abrupt reduction of the
generalization error near $\alpha=0.8$. This might be a sign of a
transition to (almost) perfect model selection, such that for
$\alpha>0.8$ the student has identified the correct $J_i=0$, and its
mistakes are restricted to the actual values of those $J_i$s that are
non-zero.

\begin{figure}[htb]
        \begin{center}              
             \includegraphics[scale=0.6, angle=-90]{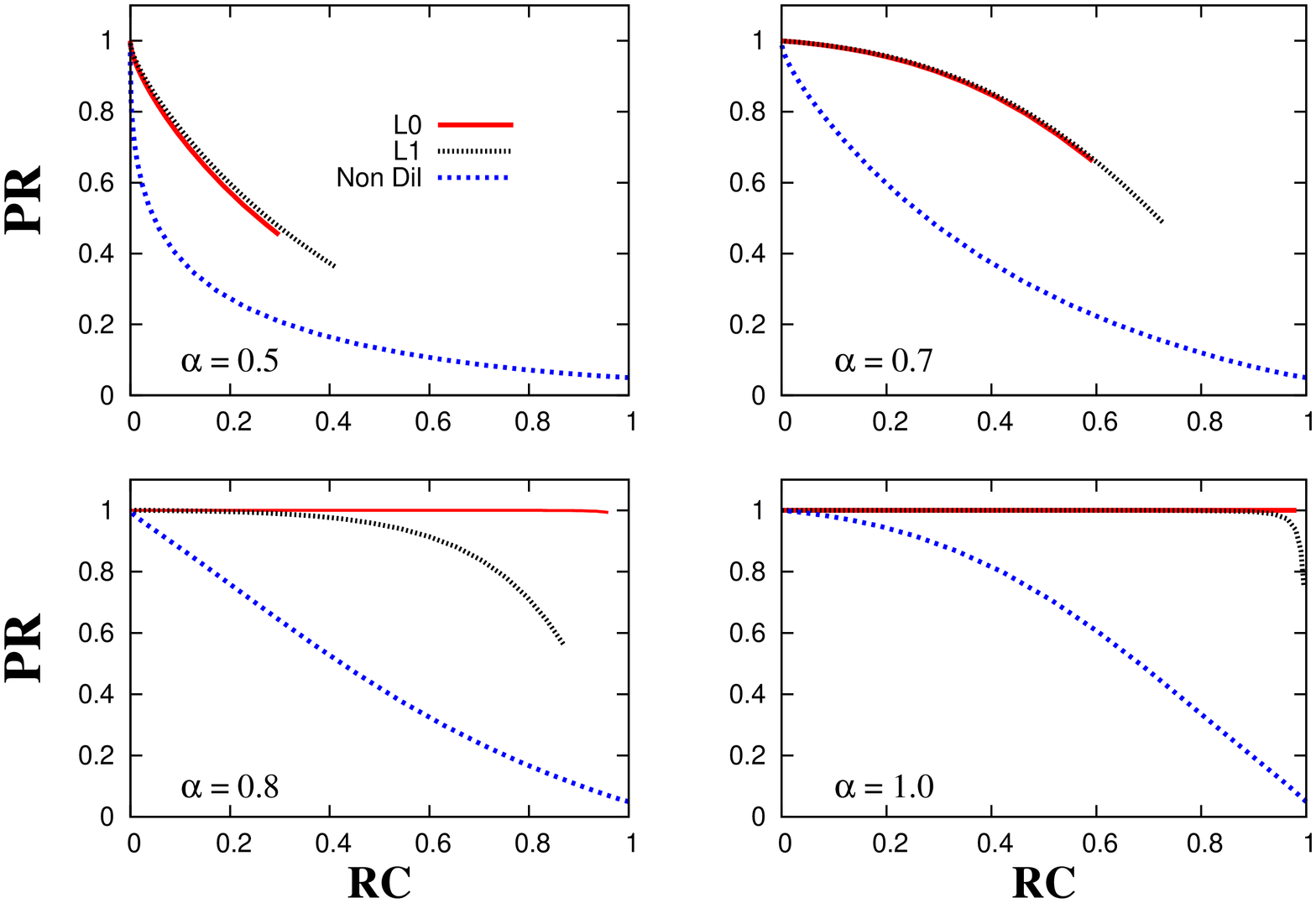}
            \end{center} 
	\caption{The PR-RC curves of the non-diluted, \Luno diluted,
          and \Lcero diluted generalization for four different values
          of $\alpha$. Both, the \Luno and \Lcero dilutions,
          outperform the non-diluted generalization. Particularly, the
          \Lcero dilution moves suddenly to almost perfect model
          selection near $\alpha=0.8$.}
\label{fig:ROCmultiplot}
 \end{figure}
To compare the \Luno and \Lcero dilutions to non-diluted
generalization, we show the PR-RC curves in Fig.
\ref{fig:ROCmultiplot}, for four typical values of $\alpha$. The
curves for the diluted generalization seem to miss the right part --
but they are not. As the \Luno and \Lcero methods set a fraction of
the couplings exactly to zero, lowering the threshold $\Jth$ will
never achieve to include them as non-zero couplings, and this is why
we can not arrive at recall equal to one.

Looking to the PR-RC curves, the first obvious fact is that the
non-diluted generalization is much worse than any of the diluted
ones. The next interesting fact is that \Luno performs slightly better
than \Lcero for low values of $\alpha$, something that could be seen
also from the generalization error (Fig.  \ref{fig:generror}). Finally
the sudden change to $PR\cong1$, for $\alpha=0.8$, of the PR-RC curve
for the \Lcero dilution is saying that \Lcero fastly moves to almost
perfect model selection, as we guessed from the generalization error
curve. However, there is no critical $\alpha$, and the sudden change
is not a phase transition. This can be seen more clearly
working with less diluted teachers (for instance $\neffcero=0.1$, data
not shown). To gain some more understanding of the onset of an almost
perfect model selection it is interesting to see the distribution of
couplings, $P(J)$, which are shown in appendices \ref{ap:hlimit} and
\ref{ap:plimit}.

\section{The memorization limit}
\label{sec:memorization}

In the calculations presented so far, the noise $\eta^\mu$ affecting
the output $y^\mu$ in Eq. \eq{eq:sigmacero}, was neglected by setting
its variance to $\gamma^2=0$. By doing so, we guaranteed that for any
$\alpha$, there is always at least one zero-cost solution for the
student, namely $\vecJ=\vecjcero$. Let us now study the opposite
extreme case where the noise is extremely large. In that
case, the output function is given by
\begin{equation}
y^\mu=\sigma^0(\vec{x})= \sig( \eta^\mu) \ , \nonumber
\end{equation}
i.e. the patterns are randomly classified by $y^\mu=\pm 1$. The
teacher's couplings $J_i^0$ become irrelevant, and the student will
try to learn (generalize) a non-existing hidden relation. This limit
is equivalent to the well-studied memorization problem of random
input-output relation.  It is a classical result \cite{Gardner88} that
for $\alpha > 2$ the student will fail to correctly classify all
patterns, while for $\alpha < 2$ the student can find a solution of
zero energy. In the latter case, the student vector $\vecJ$ reproduces
correctly the relation between the $M$ patterns $\vecxmu$ and the
corresponding labels $y^\mu$ by $y^\mu=\sig(\vecJ\cdot \vecxmu)$.  The
student was capable of memorizing the labeling of the input patterns.

Therefore memorization can be studied as the noise-dominated limit of
generalization. By computing the limit $\gamma\to \infty$ in the fixed
point equations for $\hatq$ and $\hatr$ \eq{eq:fixedpoint}, we found
that $\hatr=0$ while
\begin{equation}
 \hatq= \frac{\alpha}{\sqrt{2 \pi q (1-q)}} \int \DNx
 \frac{x}{H(\sqrt{q}x)} \label{eq:hatq}
\end{equation}
The expectation value of a function $g(x,J)$ becomes
\begin{equation}
\mathbf{E}[g(x,J)]= \int \DNx \frac{ \int \ud J\: g(x,J)
  e^{-(\frac{\hatq}{2}-\lambda)J^2 -h \norm{J} - \sqrt{\hatq} x)
    J}}{\int \ud J e^{-(\frac{\hatq}{2}-\lambda)J^2 -h \norm{J} -
    \sqrt{\hatq} x) J}} \label{eq:memPdil}
\end{equation}
In particular we have $r=\mathbf{E}[J^0 J]=0$, meaning that the
overlap between student and teacher is zero, which is an obvious
consequence of the large-noise limit. So, the set of variational
parameters describing our problem reduces to $q, \hatq$ and $\lambda$,
and the fixed-point equations are \eq{eq:hatq} and
\[ q = 1+ \frac{1}{\sqrt{\hatq}} \mathbf{E}[x J] \qquad  
1 = \mathbf{E}[ J^2] \ .
\]
The generalization error and the precision-{\it vs.}-recall curve are
meaningless in this context. However, we can still check the
efficiency of the \Luno and \Lcero memorizations in using as few as
possible non-zero couplings to memorize a set of patterns. There is a
first trivial conclusion, coming from the already stated fact that a
continuous perceptron is capable of memorizing without error until
$\alpha=2$. This means that a perceptron with $N$ couplings $J_i$ can
remember the classification of $M=2 N$ patterns. It follows directly
from this that if $\alpha<2$ patterns are given, we can set to zero
any fraction $1-\alpha/2$ of the couplings, and still be capable of
memorizing without error with the remaining $\alpha/2$ couplings. We
are interested in how much more dilution can be obtained by the
introduction of a dilution term in the Hamiltonian. Note that if,
instead of setting to zero a random group of $(1-\alpha/2)N$
couplings, we optimize their selection, we can go far below the
trivial $\neff=\alpha/2$ dilution.
 
\begin{figure}[htb]
        \begin{center}              
             \includegraphics[scale=0.8]{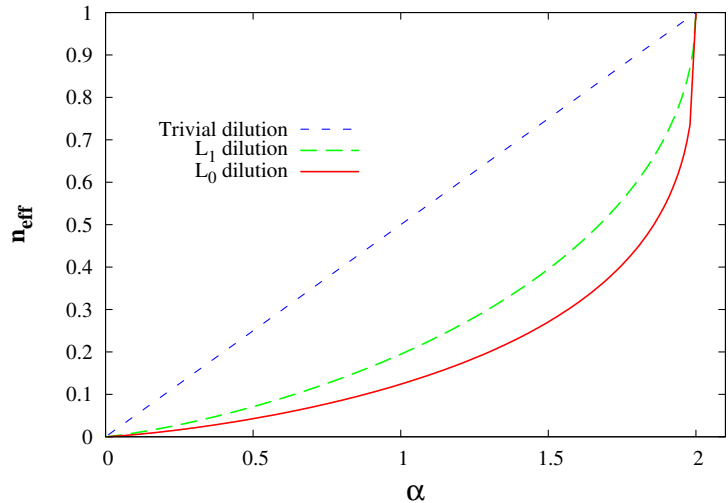}
            \end{center} 
	\caption{The maximum dilution achieved when using the \Luno
          and \Lcero regularizations in the Memorization problem. The
          trivial dilution $\neff=\alpha/2$ is outperformed by either
          \Lcero and \Luno dilutions, and \Lcero is the more efficient
          of all. For $\alpha>2$ the student fails to memorize all
          patterns and there is no student solution of zero energy. }
\label{fig:neff}
 \end{figure}

We can solve the fixed-point equations in the limit of large dilution
fields $h\to \infty$. Once again the solution space reduces to only
one solution $q\to 1$, so there are some divergences in the
equations. The scaling behavior of the variational parameters is the
following
\begin{eqnarray*}
  (1-q) &\simeq& \frac{Q}{ h} \\ \hatq&\simeq&\hatQ h^2
  \\ \frac{\hatq}{2}-\lambda &\simeq& \frac{K}{2} h
\end{eqnarray*}
Using this scaling, the expectation values are given by {\small\[
  \mathbf{E}[g(x,J)]=\int \DNx \left\{\begin{array}{lll} g(x,0) & \:\:
  & x^2<\frac{1}{\hatQ} \\ g(x,\frac{\sqrt{\hatQ} x -\sig(x)}{ K}) &
  \:\: & \mbox{ otherwise}
\end{array} \right.
\]}
for the \Luno case, and {\small\[ \mathbf{E}[g(x,J)]=\int \DNx
  \left\{\begin{array}{lll} g(x,0) & \:\: & x^2<\frac{2 K}{\hatQ}
  \\ g(x,\frac{\sqrt{\hatQ} x}{ K}) & \:\: & \mbox{ otherwise}
\end{array} \right.
\]}
for the \Lcero case.

Solving numerically the corresponding fixed point equations, we can
compute the dilution achieved by each method
\begin{eqnarray*}
n_{\mbox{eff}} &=&1- D[\delta_{J,0}]=\left\{\begin{array}{ll} 2
H\left(\hatQ^{-1/2}\right) &\qquad \mbox{L}_1\: \mbox{Norm}\\ 2
H\left(\sqrt{2 K} \hatQ^{-1/2} \right)&\qquad \mbox{L}_0\: \mbox{Norm}
\end{array}\right.
\end{eqnarray*}
The resulting functions $\neff(\alpha)$ are show in Figure
\ref{fig:neff}. It shows that both \Luno and \Lcero achieve a much
stronger dilution than the trivial random one. As in the
generalization case, the \Luno regularization works worse than \Lcero,
the reason being that it penalizes large $J$ values. The dilution
achieved in memorization for the \Luno and \Lcero dilutions is always
above the corresponding generalization curves in
Fig. \ref{fig:Genneff005}. Although not shown in either figures, we
checked that near $\alpha=0$ the corresponding curves coincide, as
there is no difference between learning and memorizing when too few
training data are given.

\section{Conclusions}
\label{sec:conclusiones}

In this paper, we have presented an analytical replica computation on
the generalization properties of a sparse continuous perceptron.
Dilution has been achieved in different ways: First, it can be imposed
naively by using non-diluted inference, followed by deleting all those
couplings which are below some threshold value. Second, it can be
achieved by introducing a dilution field which is coupled to the
$L_p$-norm of the coupling vector, penalizing thereby vectors of high
norm. For $p\leq 1$, the cusp-like singularity of the $L_p$-norm in
zero forces a finite fraction of all couplings to be exactly zero.  We
have studied in particular two special cases: (i) $p=1$ is a popular
choice in convex optimization since it is the only value of $p$ which
corresponds both to a convex penalty function and dilution.  (ii)
$p=0$ achieves optimal dilution since it penalizes equally all
non-zero couplings independently on their actual value, but due to the
non-convex character of this penalty, it easily leads to computational
intractability.

As a first finding, we see that both $L_p$ schemes work fundamentally
better than the naive scheme, both in the questions of model selection
(i.e. for the identification of topological properties of the
data-generating perceptron given by its non-zero couplings) and in the
generalization ability. For a very small or a very large amount of
training data, $L_0$ and $L_1$ achieve very comparable results. We
find, however, an intermediate regime where $L_0$ suddenly improves
its performance toward almost perfectly model selection, whereas $L_1$
dilution shows a more gradual increase in performance. This is very
interesting since this regime is found for relatively small data sets,
and in many current inference tasks (e.g. in computational biology)
the quantity of data is the major limiting factor for the
computational extraction of information. It might be in this parameter
region, where statistical-physics based algorithms like the ones
presented in \cite{Kaba03,Uda,Kabashima2007,Japan,Martin08,Tria09} may
outperform methods based on convex optimization proposed in
\cite{LASSO}.

These analytic results call for efficient algorithms in real case
studies. At odds with the linear-regression case with \Luno norm, in
the case of a continuous perceptron, a simple gradient descent
strategy does not work due to the presence of a zero-mode in the
energetic term Eq.~(\ref{eq:E}) ($E(\vecJ) = E(c\vecJ)$ for every
scalar $c>0$). The zero-mode has been removed in the computation by
fixing the modulus of the classification vector
($\vecJ\cdot\vecJ=N$). Unfortunately this spherical constraint breaks
the convexity of the problem and it is not clear if there are more
ingenious ways for removing the zero-mode that could work, at least in
the \Luno norm case. Another possibility that we are planning to
follow is that of considering variational approximation schemes like
belief propagation for continuous perceptrons \cite{Kabashima2007,
Martin08, Tria09}, which are able to overcome also the problem of the
non-convexity of the $L_0$ norm.

During the preparation of this manuscript, a related study on the
efficiency of \Lp dilution in systems of linear equations was posted
online \cite{kaba09}. Also there, the relative importance of $L_0$ and
$L_1$ dilution was studied, with conclusions which are highly
compatible to ours.

% If you have acknowledgments, this puts in the proper section head.
\begin{acknowledgments}
A.L. and M.W. acknowledge support by the EC-founded STREP GENNETEC
(``Genetics Networks: emergence and copmlexity'').
\end{acknowledgments}

\appendix

\section{Replica calculation details}
\label{ap:Replica}
The calculation of $\overline{Z^n}$ in Eq. \eq{eq:Zn} is done by the
introduction of an overlap matrix $Q_{a,b}$ using constraints
\[\delta(N Q_{a,b} - \sum_i J_i^a J_i^b)
\]
for any $0\leq a\leq b\leq n$. As there is a symmetry in the replica
indices $Q_{a,b}=Q_{b,a}$, only the half of the matrix is needed. The
value $a=0$ refers to the teacher, while $a=1\ldots n$ to the $n$-fold
replicated student. Among these constraints, there are some that are
particular. For instance the term $Q_{0,0}$ is the variance of the
teacher, and it should be equal to the variance $t$ of the teachers
distribution \eq{eq:rhoJ}. Similarly, the $n$ terms $Q_{a,a}$ are set
to $1$, in order to impose the spherical constraint on the student,
since the energy \eq{eq:E} is invariant to elongations of the student
vector.

Using Fourier representation of the Dirac-deltas, the replicated
partition function is {\small\begin{eqnarray} \overline{Z^n}=\int
  \frac{\ud Q_{a,b} \ud\hat{Q}_{a,b}\ud^n \lambda^a}{(2 \pi)^{3/2}
    N^{-1}} \exp\left(i N \sum_{a< b} Q_{a,b} \hat{Q}_{a,b} +i N
  \sum_{a>0} \lambda_a +i N t \lambda_0\right) \nonumber \\ \left(
  \int \ud^n J^a \rho(J^0) \exp(-h \sum_a^n \norm{J^a} -i \sum_{a\leq
    b}J^a \hat{Q}_{a,b} J^b ) \right)^N \label{eq:ZnQ} \\ \left(\int
  \mathrm{D_\gamma}\eta \frac{\ud^n X^a \ud^n\hat{X}^a}{2 \pi}
  \exp(-\B \sum_a^n \theta\left(-(X^0+\eta) X^a\right) +i\sum_{a}X^a
  \hat{X}^a-\frac{1}{2}\sum_{a,b} \hat{X}^a Q_{a,b} \hat{X}^b)
  \right)^M \nonumber
\end{eqnarray}}
where $\hat{Q}_{a,b}$ are the conjugated parameters in the Fourier
representation of the deltas. In particular, $\lambda^0$ and
$\lambda^a$ are the one corresponding to the teacher variance and the
spherical constraint. To save some space, we used the short-hand
notation $\ud^n A^a$ as a substitute for $\prod_{a=0}^n \ud A^a$, and
$\ud Q_{a,b}$ for the differential of all the terms in the overlap
matrix.

The next step in the replica calculation is to assume a structure for
the overlap matrix. In the replica-symmetric case, the overlap matrix
and its Fourier counterpart have the structure (exemplified for $n=3$)
{\small\begin{equation} Q_{a,b}=\left(\begin{array}{cccc} t & & &\\ r
    & 1 & &\\ r & q & 1 &\\ r & q & q & 1
\end{array}\right) \quad -i \hat{Q}_{a,b}=\left(\begin{array}{cccc}
\lambda^0 & & &\\ \hatr & \lambda & &\\ \hatr & \hatq & \lambda
&\\ \hatr & \hatq & \hatq & \lambda
\end{array}\right)  \label{eq:RSoverlap}
       \end{equation}
} The Fourier mode corresponding to the variance of the teacher $t$,
can be shown to be $\lambda^0=0$, while that of the spherical
constraint remains a variational parameter $\lambda=-i \lambda^a$. The
other parameters are $q$, the self overlap between two student
solutions, $r$, the overlap between an student and the teacher, and
their conjugate Fourier modes $\hatq$ and $\hatr$.

It is a standard feature of the replica trick to invert the order of
the limits, doing $N\rightarrow \infty$ first, and then $n\rightarrow
0$, profiting thereby of the saddle-point method to solve the integral
in \eq{eq:ZnQ}. Note that the last two lines in \eq{eq:ZnQ} can be
brought to the exponential by using $X=\exp\log X$. Thus the value of
the free energy $ -\B \overline{f}= \lim_{n\to 0} \lim_{N\to
  \infty}\frac{1}{N n} \log\overline{ Z^n}$ is given by extremizing
the equation
\begin{equation} -\B \overline{f}=-r\hatr+\frac{1}{2}q \hatq -\lambda 
+ G_J+\alpha\:  G_X \nonumber
\end{equation}
 with respect to the variational parameters
 $(q,r,\hatq,\hatr,\lambda)$, where we have introduced {\small
\begin{eqnarray*}
G_J & =&\int \DNx \intrho \log\int \ud J
e^{-(\frac{\hatq}{2}-\lambda)J^2 -h \norm{J} +(\hatr J^0- \sqrt{\hatq}
  x) J} \\ G_X & = & 2 \int \DNx H\left(\frac{ x r}{\sqrt{q \gamma^2+q
    t-r^2)} }\right) \log\left((e^{-\B}-1) H(-\sqrt{\frac{q}{1-q}}x)+1
\right) \nonumber
\end{eqnarray*}
} and
\begin{equation*}
H(x)=\int_x^\infty \frac{\ud y}{\sqrt{2 \pi}}e^{-y^2/2}
\end{equation*}

\section{Limit $h\rightarrow \infty$}
\label{ap:hlimit}
The scaling behavior of the parameters $q, r, \hatq, \hatr$ and
$\lambda$ in the limit $h\to \infty$
\begin{equation*}
 \begin{array}{rcllrcl}
  (1-q) &\simeq& \frac{Q}{ h} &\quad& \hatr&\simeq&\hatR h
   \\ \hatq&\simeq&\hatQ h^2 &\quad& \frac{\hatq}{2}-\lambda &\simeq&
   \frac{K}{2} h
 \end{array}
\end{equation*}
were first obtained by looking at the solutions of the fixed-point
equations for growing values of $h$, and their consistency was checked
later in the fixed-point equations. Considering this scaling, the
expectation value of a generic function $g(x,J^0,J)$ is given by
\begin{equation}
\mathbf{E}[g(x,J^0,J)]=\int \DNx \ud J^0 \rhoJ \frac{ \int \ud J\:
  g(x,J^0,J) e^{-h \left(\frac{K}{2} J^2 + \norm{J} - (\hatR J^0-
    \sqrt{\hatQ} x) J\right)}}{\int \ud J e^{-h \left(\frac{K}{2} J^2
    + \norm{J} - (\hatR J^0- \sqrt{\hatQ} x)
    J\right)}} \label{eq:apPdil}
\end{equation}
The diverging prefactor $h$ in the exponentials forces the main
contribution to the $J$-integration to come from the largest value of
the exponent (saddle-point approximation):
\begin{equation}
J^*_p=\mathop{{\argmin}\vphantom{\sim}}\limits_{\displaystyle
  _{\mathbf J}} \left(\frac{K}{2} J^2 + \norm{J} - (\hatR J^0-
\sqrt{\hatQ} x) J\right) \label{eq:Jmin}
\end{equation}
In the case of the \Luno norm ($\|J\|_1=|J|$) the solution of the
previous equation is given by
\begin{equation}
 J=\left\{\begin{array}{lll} 0 & \:\: & |\hatR J^0 -\sqrt{\hatQ} x|<1
 \\ \frac{\hatR J^0 -\sqrt{\hatQ} x -\sig(\hatR J^0 -\sqrt{\hatQ} x)}{
   K} & \:\: & \mbox{ otherwise}
\end{array} \right.
\end{equation}
The expectation value is thus {\small\[ \mathbf{E}[g(x,J^0,J)]=\int
  \DNx \intrho \left\{\begin{array}{lll} g(x,J^0,0) & \:\: & |\hatR
  J^0 -\sqrt{\hatQ} x|<1 \\ g(x,J^0,\frac{\hatR J^0 -\sqrt{\hatQ} x
    -\sig(\hatR J^0 -\sqrt{\hatQ} x)}{ K}) & \:\: & \mbox{ otherwise}
\end{array} \right.
\]}
\begin{figure}[htb]
        \begin{center}              
             \includegraphics[scale=0.6, angle=0]{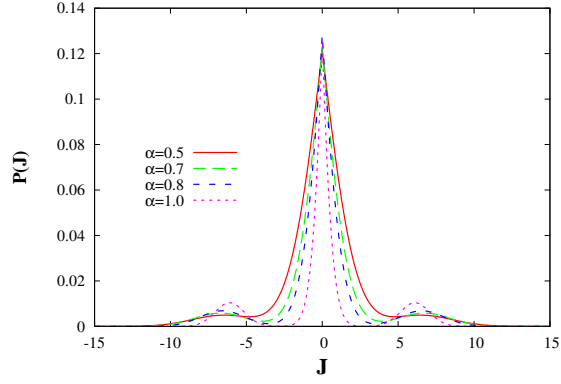}
            \end{center} 
	\caption{The distribution $P(J)$ of the student couplings for
          the four values of $\alpha$ used in the Precision-Recall
          curves of Fig. \ref{fig:ROCmultiplot}. }
\label{fig:PJL1}
 \end{figure}

The probability distribution of the students couplings $P(J)$ can be
obtained as $P(J')=\mathbf{E}[\delta(J-J')]$ resulting in
\[
P(J)=(1-\neff) \:\delta(J) + \frac{K}{\sqrt{\hatQ}\sqrt{2 \pi}}\intrho
e^{-\frac{(\hatR J^0- \sig J -K J)^2}{2 \hatQ}}
\]
where $\neff=1 - \int \DNx \intrho \Theta[|\hatR J^0 -\sqrt{\hatQ}
  x|-1]$. The continuous part of this distribution is shown in
Fig. \ref{fig:PJL1} for the same four values of $\alpha$ for which the
Precision-Recall curves were shown in Fig.~\ref{fig:ROCmultiplot}. We
can see that for growing values of $\alpha$, the distribution $P(J)$
is more concentrated around the discrete values of $J$, and the amount
of couplings that are small but not zero, reduces continuously. This
explains the high performance in model selection of the \Luno dilution
for $\alpha>1.0$.

Note that the calculation of the smallest value of the exponent in
\eq{eq:apPdil} is particularly simple for \Lcero and \Luno. Other
values of $p$ may require a numerical solution. It is simple to see
that in the $p>1$ case no dilution is obtained.

\section{\Lcero as the $p\to 0$ limit}
\label{ap:plimit}
The \Lcero dilution corresponds to a term $\sum_i (1-\delta_{J_i})$ in
the Hamiltonian \eq{eq:Ham}. However, the Kronecker delta is zero for
all non-zero arguments, with an isolated and finite discontinuity in
the origin. This single-point discontinuity is irrelevant in the
integration over continuous $J$s in the partition function as well as
in \eq{eq:apPdil}. Therefore using the \Lcero dilution from the
beginning gives the same results as the non-diluted case
$h=0$. Nevertheless, we can still interpret the \Lcero norm as the
$p\to+0$ limit of the \Lp norm.
\begin{figure}[htb]
        \begin{center}              
             \includegraphics[scale=0.6, angle=0]{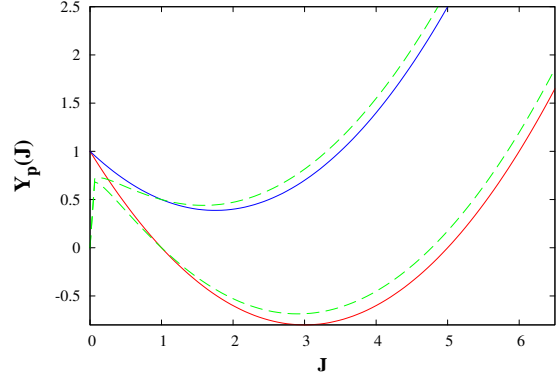}
            \end{center} 
	\caption{The $p=0$ and $p>0$ cases of the function
          $Y_p(J)=\frac{K}{2} J^2 + \norm{J} - (\hatR J^0-
          \sqrt{\hatQ} x) J$ in the two characteristic situations
          where $J^*_p=0$ and $J^*_p>0$. The closer $p$ to zero, the
          closer the function $Y_p(J)$ is to $J_0(J)$.}
\label{fig:Lpcuadratic}
 \end{figure}

For general $p> 0$ there is no explicit solution for
Eq. \eq{eq:Jmin}. We will argue that the limit $p\to 0$ of such
solutions is exactly the solution of
\[
J^*_0=\mathop{{\argmin}\vphantom{\sim}}\limits_{\displaystyle
  _{\mathbf J}}\left(\frac{K}{2} J^2 + (1-\delta_J) - (\hatR J^0-
\sqrt{\hatQ} x) J\right)
\]
just as if we would have introduced the \Lcero norm from the
beginning, and taken naively the saddle point including the isolated
singularity. There are two candidate values for $J^*_0$, one is $0$
and the other one is the zero-derivative point of the quadratic
function $J^*_0=\frac{\hatR J^0-\sqrt{\hatQ}}{T}$. The latter will be
the actual solution if and only if
\[\frac{(\hatR J^0-\sqrt{\hatQ})^2}{2 T}<1
\]
If the opposite inequality is satisfied, the solution is
$J^*_0=0$. Both situations are shown in Figure
\ref{fig:Lpcuadratic}. The function $|J|^p$ tends to $1$ as $p\to 0$
for all $J\neq 0$, so we have also $J^*_p\to J^*_0$ whenever $
\frac{(\hatR J^0-\sqrt{\hatQ})^2}{2 T}\neq1$. The point where the
equality holds corresponds to the neglectable case when the value in
$J=0$ is exactly equal to that in the point of zero derivative. We
conclude that except for this single point, $J^*_p\to J^*_0$ as $p\to
0$, and therefore we can replace the \Lcero norm directly into the
steepest descend condition to obtain the $p\to 0$ result.
\begin{figure}[htb]
        \begin{center}              
             \includegraphics[scale=0.6, angle=0]{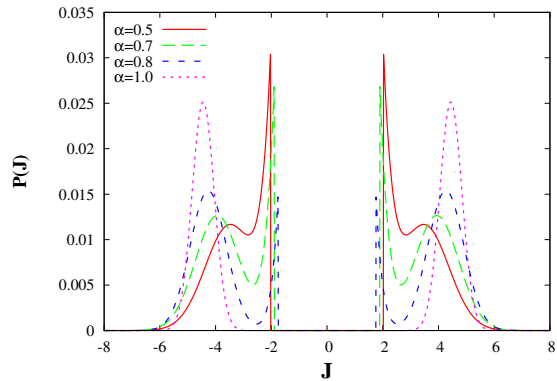}
            \end{center} 
	\caption{The distribution $P(J)$ of the student couplings for
          the four values of $\alpha$ used in the Precision-Recall
          curves of Fig. \ref{fig:ROCmultiplot}. Comparing also this
          result for \Lcero with the one for \Luno in
          Fig. \ref{fig:PJL1}, we can understand the difference in
          their performance.}
\label{fig:PJL0}
 \end{figure}

Repeating the steps shown in appendix \ref{ap:hlimit}, a similar
computation for the \Lcero dilution gives the expectation value
reported in Eq. \eq{eq:ExpectedL0}, and the following probability
distribution for the student couplings
\[
P(J)=(1-\neff) \:\delta(J) + \frac{K}{\sqrt{\hatQ}\sqrt{2 \pi}}
\intrho e^{-\frac{(\hatR J^0-K J)^2}{2 \hatQ}}
\Theta(|J|-\sqrt{\frac{2}{T}})
\]
This distribution is shown in Fig. \ref{fig:PJL0} for the same four
values of $\alpha$ for which the Precision-Recall curves were shown in
\ref{fig:ROCmultiplot}. Note that the main difference between this
distribution and the corresponding to the \Luno dilution
Fig. \ref{fig:PJL1} is the presence of the $\Theta(\cdot)$ function in
the former. When the Gaussians of the continuous part of the
distribution have a standard deviation smaller than the gap in the
$\Theta$ function, the presence of False Positives corresponding to
the Gaussian around $J^0=0$ is suppressed by the $\Theta$ function, and
this is the reason why we observe such a good performance in model
selection for $\alpha>0.8$ in Fig. \ref{fig:ROCmultiplot}.

\bibliography{biblio}
\end{document}